# Efficiency-boosted semiconductor optical amplifiers via mode-division multiplexing

Yi Wang[1,*], Yihui Wei[1], Victor Dolores-Calzadilla[1,2], Daoxin Dai[3,4], Kevin Williams[1], Meint Smit[1], and Yuqing Jiao[1]

[1] Eindhoven Hendrik Casimir Institute (EHCI), Eindhoven University of Technology, Eindhoven, 5600MB, the Netherlands

[2] Photonic Integration Technology Center (PITC), Eindhoven University of Technology, Eindhoven, 5600MB, the Netherlands

[3] State Key Laboratory for Modern Optical Instrumentation, Center for Optical & Electromagnetic Research, College of Optical Science and Engineering, International Research Center for Advanced Photonics, Zhejiang University, Zijingang Campus, Hangzhou 310058, China.

[4] Jiaxing Key Laboratory of Photonic Sensing & Intelligent Imaging, Intelligent Optics & Photonics Research Center, Jiaxing Research Institute, Zhejiang University, Jiaxing 314000, China.

*Corresponding author: y.wang10@tue.nl

# Emails of all authors:

Yi Wang:

y.wang10@tue.nl

Yihui wei:

y.wei1@tue.nl

Victor Dolores-Calzadilla:

v.calzadilla@tue.nl

Daoxin Dai:

dxdai@zju.edu.cn

Kevin Williams:

k.a.williams@tue.nl

Meint Smit:

m.k.smit@tue.nl

Yuqing Jiao:

y.jiao@tue.nl

**Semiconductor optical amplifiers (SOA) are a fundamental building block for many photonic systems. However, their power inefficiency has been setting back operational cost reduction, and the resulting thermal losses constrain miniaturization, and the realization of more complex photonic functions such as large-scale switches and optical phased arrays. In this work, we demonstrate significant gain and efficiency enhancement using an extra degree of freedom of light – the mode space. This is done without changing the SOA's material design, and therefore high versatility and compatibility can be obtained. Light is multiplexed in different guided modes and is reinjected into the same gain section twice without introducing resonance, doubling the interaction length in a broadband manner. Up to 87% higher gain and 300% higher wall-plug efficiency are obtained in a double-pass SOA compared to a conventional single-pass SOA, at the same operating current, in the wavelength range of 1560 – 1580 nm.**

## Introduction

The pursuit of high energy efficiency has fueled major advances in photonic microchips, as it promises denser integration, lower costs, and reduced environmental impact of advanced technologies[1–5]. The semiconductor optical amplifier (SOA), a key device at the interface between electronics and photonics, is unfortunately one of the most power-hungry and inefficient components in photonic integrated circuits (PIC), especially when weak signals are to be amplified. This is due to the fact that the strength of stimulated emission depends linearly on the photon density[6], which increases exponentially along the propagation direction. The mismatch between photon density and available gain can therefore leads to high levels of amplified spontaneous emission when signals are low, contributing to increased noise rather than signal amplification. The carrier density is uneven along the propagation direction, and the carriers are not efficiently utilized for stimulated emission before the photon density is sufficiently high.

Several approaches have been taken to improve the SOA's energy efficiency, including increased optical confinement[7–9], deeper quantum wells (QW)[10–12], better heat sinking[13–17], higher injection efficiency[18–20], etc. Slow light has also been employed for SOA efficiency enhancement[21], but only optical pumping has been demonstrated. To the best of our knowledge, all these existing works focus on improving the SOA at the device level, so they are specific to a certain technology platform. They are all single-pass devices, so they still suffer from the problem of spatially-inhomogeneous and inefficient carrier utilization.

Here, we propose a novel direction of SOA efficiency enhancement at the circuit level, leveraging on-chip mode-division-multiplexing (MDM)[22–24]. In our approach, light gets reinjected back and forth through the SOA, reusing the same gain material in both directions. Therefore, the interaction length is multiplied and the available carriers are utilized in a more homogeneous way. Resonance is suppressed since each pass is spatially multiplexed in separate modes. This is a universal method which in principle can be applied to any active-passive platforms without changing the active layer structures. Efficiency enhancement using MDMs has been demonstrated for linear devices such as thermo-optic[25] or electro-optic[26–28] modulators, but it has never been done for SOAs which are highly nonlinear: the gain is naturally exponential along the propagation in absence of gain saturation. Therefore, in SOAs, higher efficiency gain can be expected with the same number of passes, or fewer passes may be needed to achieve a substantial efficiency boost.

The challenge of integrating MDMs with active devices mainly lies in their distinct waveguide cross-sections. Low-loss and broadband MDM (de)multiplexers normally require rectangular or ridge waveguides[22,29] for efficient evanescent coupling, but typical SOAs are realized in micron-scale vertical layer stacks[30]. Heterogeneous integration of SOAs onto high-confinement platforms using bonding or transfer printing could enable such co-integration, but fabrication tolerances must be considered. In this demonstration, the challenge is tackled by the InP membrane on silicon (IMOS) technology[31,32], where the functional layers are precisely defined by epitaxy, and nanophotonic passive circuits and SOAs can be monolithically integrated at wafer scale.

In this paper, we demonstrate, for the first time, significant SOA efficiency enhancement through a double-pass (DP) gain section, leveraging on-chip MDM. The design, fabrication, and characterization results are subsequently presented in the following sections.

**Results**

**Circuit design.** The generic circuit architecture of the DP-SOA is shown in Fig. 1**a**. The light path is as follows: Input fiber grating coupler (FGC, $TE_0$) – MDM (de)multiplexer ($TE_0$ passing through) – SOA ($TE_0$ gain) – MDM (de)multiplexer ($TE_0$ passing through) – U-bend ($TE_0$) – MDM (de)multiplexer ($TE_0$ converted to $TE_1$) – SOA ($TE_1$ gain) – MDM (de)multiplexer ($TE_1$ converted back to $TE_0$) – output FGC ($TE_0$). Although a high-order mode is used inside the circuit, fiber interfacing is done only using the fundamental mode. As seen in Fig. 1**b**, by breaking apart the U-bend and connecting to FGCs, an alternative MDM 2 × 2 SOA is obtained, so the $TE_0$ and $TE_1$ gain channels can be independently accessed and characterized. This single-pass $TE_0$ channel can be used as a benchmark to quantify the gain performance of the DP-SOA.

The DP-SOA is designed on the IMOS platform, on which all photonic devices are realized in a thin InP-based epilayer adhesively bonded on Si. The SOA gain section is based on a regrowth-free twin-guide scheme[31,32]. The active functional layers including the n- and p-InP claddings, the InGaAsP Q1.25 core containing a compressively-strained 4-multi-quantum-well (MQW), and the metal contact layers are epitaxially grown on top of the 300 nm-thick i-InP layer, in which the passive waveguide and other photonic circuitries are fabricated. Halfway fabrication, the whole layer stack is flipped and buried in the benzo-cyclobutene (BCB) bonding layer, as seen in Fig. 1**c**. The whole wafer is also planarized and buried in polyimide (PI). The Q1.25 layer immediately under the i-InP layer is a wet etch-stop layer, which ensures an accurate etch depth for the SOA ridge and tapers without damaging the passive waveguide (see Methods). The width of the SOA is designed to be 2.2 μm, supporting both the $TE_0$ and $TE_1$ mode in the core region. Computed modal fields are shown in Supplementary Fig. S1. The confinement factors in the 4-MQW are calculated to be 4.31% and 4.29% for $TE_0$ and $TE_1$, respectively.

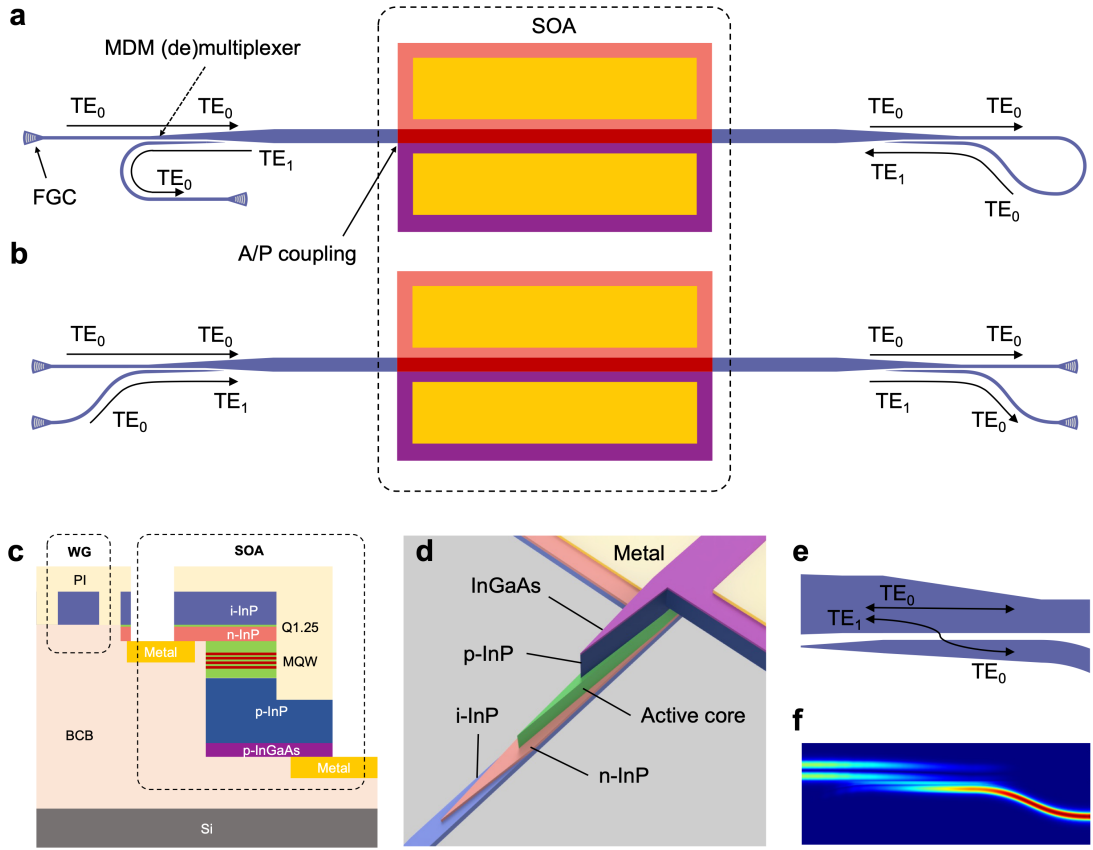

**Figure 1: Schematics of the MDM-SOA circuits (not-to-scale). a:** Top view of the DP-SOA. **b:** Top view of the MDM 2 × 2 SOA. **c:** Cross-sectional structure of the passive waveguide and the SOA. The color scheme is the same as in **a**. **d:** 3-D illustration of the SOA active/passive taper, looking from the substrate side. **e:** Zoom-in schematics of the MDM (de)multiplexer. **f:** Mode coupling between $TE_0$ and $TE_1$ through the MDM (de)multiplexer in **e**.

The active/passive (A/P) coupling between the SOA and the passive waveguide is done via an evanescent taper, as seen in Fig. 1**d**. In this taper, by separately tapering down the thick p-InP, eliminating vertical high-order modes in the p-cladding, the unwanted mode conversions are suppressed (Supplementary Fig. S2). The p-cladding, core, and n-cladding taper lengths are 10 μm, 35 μm, and 20 μm, respectively. The p-cladding and core taper both start at the gain section, and therefore the total taper length is 55 μm. The tip widths of the tapers are set to 200 nm for fabrication robustness[33]. According to eigenmode expansion (EME) simulations, a high 90% transmission (Supplementary Fig. S3) is achieved for both the $TE_0$ and $TE_1$ modes in this taper.

Fig. 1**e** shows the zoom-in schematics of the dual-core adiabatic tapers as an MDM (de)multiplexer. This MDM (de)multiplexer contains a multimode bus connecting directly to the SOA, and a $TE_0$ access waveguide for adiabatic coupling from/to the $TE_1$ mode in the bus waveguide. The gap between the waveguides is designed to be 100 nm, allowing for simultaneously good manufacturability and a small footprint (50 μm coupling length). Detailed structural parameters and performance of the MDM (de)multiplexer can be found in Supplementary Note 3. Fig. 1**f** shows the light propagation through this MDM (de)multiplexer when $TE_0$ light of 1550 nm wavelength is launched in the access waveguide. This design is demonstrated to have high fabrication tolerance on waveguide width (±25 nm), low insertion loss (<1 dB), and broad optical bandwidth (90 nm)[22].

**Fabrication.** The fabrication is done on both sides of the InP membrane (before and after bonding to Si, see Methods for details) to realize the twin-guide S-shaped cross-sectional structure. The bonding layer thickness is 2 μm, enabling high fabrication tolerance to surface topologies and particles. The passive waveguides, MDM (de)multiplexers, and FGCs are fabricated after bonding and are therefore on top of the bonding layer. A microscope image of the fabricated DP-SOA is shown in Fig. 2**a**. The zoom-in scanning electron microscope (SEM) images of the active-passive taper before bonding and the coupling region of the MDM (de)multiplexer are shown as insets of the figure. Since e-beam lithography (EBL) is used for pattern definition, high overlay accuracy (~ 20 nm) is achieved for all mask layers.

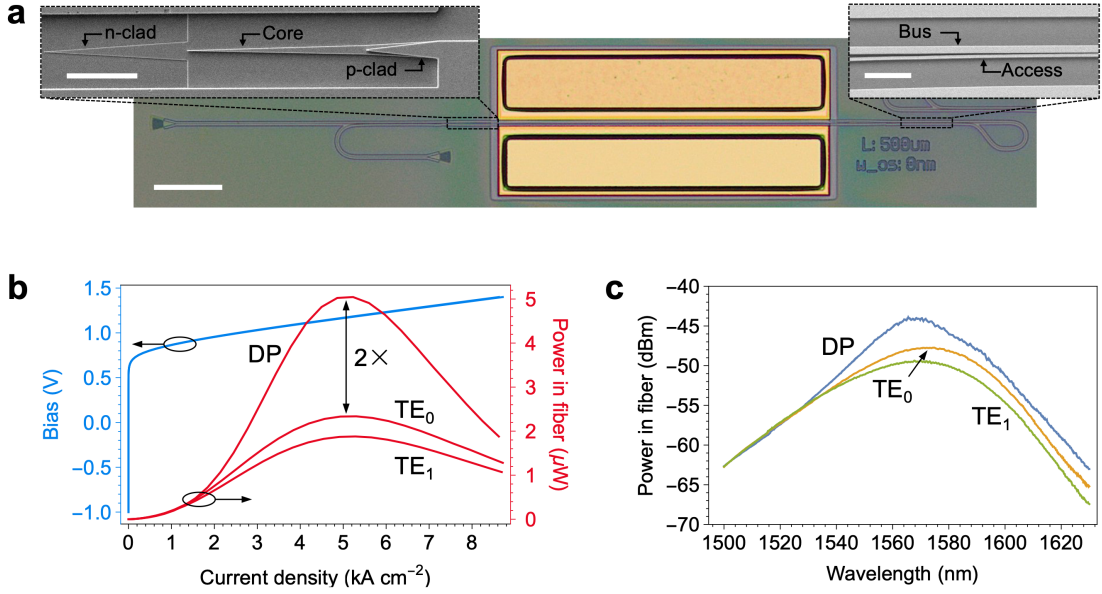

**Figure 2: ASE enhancement of the DP-SOA. a:** Microscope image of the fabricated DP-SOA (scale bar: 100 μm). The insets are zoom-in SEM images. Top left: SOA active-passive taper before bonding (scale bar: 10 μm). Here the i-InP waveguide is not yet fabricated. Top right: MDM (de)multiplexer fabricated after bonding. **b:** LIV curves of the DP-SOA and the reference $2 \times 2$ SOA of the same length. The received optical power for each SOA channel is labelled in the figure. **c:** ASE spectra of the DP-SOA and the reference $2 \times 2$ SOA of the same length at a current density of 4.1 kA cm$^{-2}$.

**MDM-enhanced amplified spontaneous emission (ASE).** The DP-SOA and a nearby $2 \times 2$ SOA with the same gain section dimensions ($2.2 \times 500$ μm) are measured to characterize the gain enhancement. The wafer stage is kept at 10°C, and cleaved single mode fibers (SMF) are used for optical interfacing with the FGCs at an emission angle of 80° with respect to the waveguides. First, the LIV curves are obtained using a current source and a broadband optical power meter for both channels of the $2 \times 2$ SOA and the DP-SOA. As seen in Fig. 2**b**, owing to the high fabrication uniformity, the IV curves of the $2 \times 2$ SOA and the DP-SOA overlap well with each other, and they both have a series resistance of ~7 Ω. As can be seen, due to an extra pass in the gain section, $2 \times$ higher ASE power is obtained at the same current density for the DP-SOA compared to the TE$_0$ channel of the $2 \times 2$ SOA (i.e., single pass, SP). Since the TE$_0$ gain channel does not involve mode coupling over the MDM (de)multiplexer, the excess loss is negligible[22]. Therefore, this channel can be used to benchmark the performance of the DP-SOA. The TE$_1$ ASE is lower than that of the TE$_0$, which is due to the $TE_1$ channel having: 1) extra loss from the MDM (de)multiplexer (Supplementary Fig. S5). 2) a slightly lower confinement factor and higher sidewall scattering loss in the gain section. Optical power saturation occurs at around 5 kA cm$^{-2}$, and the optical power decreases beyond this point. The ASE spectra collected by an optical spectrum analyzer (OSA) at the current density of 4.1 kA cm$^{-2}$ for the DP-SOA and the 2 × 2 SOA are shown in Fig. 2**c**. Near-smooth (<0.5 dB) spectra are obtained for both the $TE_0$ and $TE_1$ channels, indicating the high mode conversion efficiency and low crosstalk in the MDM (de)multiplexers. Larger fluctuations can be observed in the spectra of the DP-SOA. These fluctuations come from the residual optical feedback: ~4% reflection of the SMF facet, ~1% refection of the grating couplers, and ~1% effective reflection from the MDM (de)multiplexers.

**Gain and efficiency enhancement measured using laser input.** Since the MDM SOA circuits involve more than one mode, measurement of optical gain is done by directly comparing the output and input optical powers. To measure the small-signal gain, an external tunable laser is used as the input, whose peak wavelength is tuned from 1520 nm to 1600 nm with a step of 2 nm. The on-chip input power after the FGC is set at -13.5 dBm. The input polarization is set for maximum optical transmission under a constant current injection. Since the FGCs and MQWs are designed for TE, and TM gain is far less than TE gain, a TE input polarization is assumed. Then, at various injection current densities, optical spectra are collected by an OSA for each input-output channel combinations, for both the 2 × 2 SOA and the DP-SOA. Then, the peak powers of the spectra are extracted. In this way, the broad ASE power can be excluded from the gain measurement. Nearby (< 100 μm distance) reference waveguides are also measured at each corresponding wavelength point using the same configuration to calibrate out the FGC loss. The passive propagation loss in the access waveguides between the FGCs and the MDM (de)multiplexers are compensated, and therefore zero-net-loss levels are obtained.

Fig. 3**a** – Fig. 3**c** show the received optical spectra at the current density of 4.1 kA cm$^{-2}$, with the blue dots indicating zero-net-loss levels, for the 2 × 2 SOA with input through the $TE_0$ port, input in the $TE_1$ port, and the DP-SOA, respectively. As can be seen, positive optical gain is achieved in the wavelength range of 1540 nm – 1600 nm for all three cases, and there are no signs of lasing of the SOAs themselves. As seen in Fig. 3**a** and Fig. 3**b**, for the 2 × 2 SOA, the modal crosstalk ($TE_0$ – $TE_1$ and $TE_1$ – $TE_0$ transmission) levels are -16.2 dB – -28.6 dB and -16.1 dB – -27.8 dB in the wavelength range of 1520 nm – 1570 nm, for $TE_0$ and $TE_1$ input, respectively. This indicates that independently controllable and accessible gain channels are achieved. Generally, the crosstalk increases as the wavelength decreases, which could be due to stronger absorption and re-emission[34,35] in the gain section at shorter wavelengths.

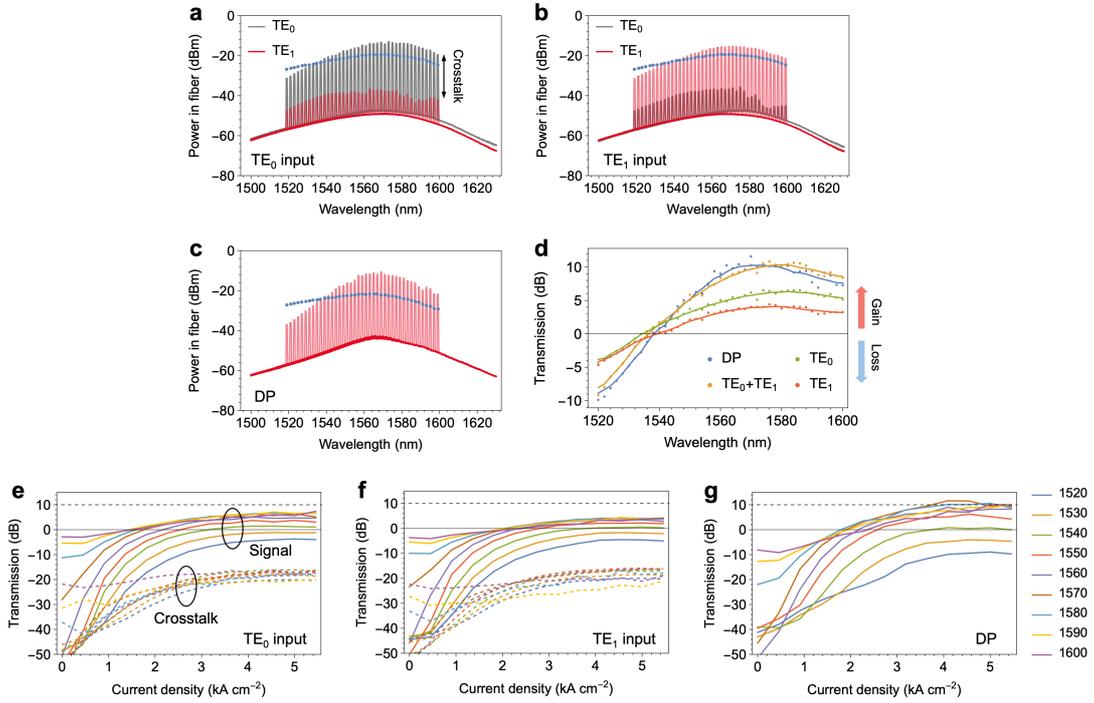

**Figure 3: Gain enhancement of the DP-SOA. a, b:** Optical spectra of the 2 × 2 SOA obtained at the current density of 4.1 kA cm$^{-2}$, using a tunable laser with -13.5 dBm on-chip input power. The spectra of different input wavelengths are overlapped, and the input and output channels are labeled on the bottom right and top right of the figures, respectively. The blue dots indicate zero-net-loss levels measured by reference waveguides. **c:** Optical spectra of the DP-SOA at the same measurement conditions as in **a** and **b**. **d:** Optical transmission through the SOAs, obtained by subtracting the reference levels from the peak powers in **a-c**. The dots are actual data points, while the solid curves are wiener-filtered to eliminate the fluctuations from FGC and fiber reflections. **e, f:** Optical gain as functions of the injection current densities at various wavelengths, calculated by subtracting the zero-net-loss powers from the obtained peak powers. The solid lines and dashed lines are transmissions from the signal channels ($TE_0$-$TE_0$ and $TE_1$-$TE_1$) and crosstalk channels

(TE$_0$-TE$_1$ and TE$_1$-TE$_0$), respectively. **g:** Optical transmission of the DP-SOA obtained using the same method as in **e** and **f**.

Fig. 3**d** shows the gain values by comparing the measured peak powers and the zero-net-loss levels. As can be seen, the gain of the DP-SOA matches well with the summation of values obtained from the separate gain channels. At shorter wavelengths where the material gain is negative, the double-pass loss also corresponds well to the total loss in the TE$_0$ and TE$_1$ channels. This indicates that light successfully travels through the gain section twice, and the zero-net-loss levels are correctly set. Compared to single-pass in TE$_0$, significant gain enhancement is achieved in the wavelength range of 1538 nm – 1600 nm for the DP-SOA. In the wavelength range of 1560 nm – 1580 nm, over 158% (58% more) optical gain is achieved, without increasing the operating current. The wall-plug efficiency (WPE, WPE = P$_{optic}$ /P$_{electric}$) can be calculated using the on-chip input and output optical power (P$_{optic}$ = P$_{out}$ – P$_{in}$) and total electrical power (P$_{electric}$ = VI). In this 20 nm bandwidth, a > 180% extra WPE is achieved. At 1570 nm in particular, 11.6 dB gain is achieved for the DP-SOA, while the gain measured for single-pass TE$_0$ is 6.2 dB. Therefore, the net gain enhancement is 5.4 dB, or 87% more than the original value, for the same operating current. The WPE for single-pass and double-pass are 0.3% and 1.2% respectively. a 300% extra WPE is achieved. At the shorter wavelengths of 1520 nm – 1536 nm, > 100% absorption enhancement can be observed, which partly attributes to the extra loss from the MDM-couplers. Fig. 3**f** – Fig. 3**g** show the gain as a function of injection current density, at various wavelengths. As can be seen, the DP-SOA and the single-pass TE$_0$ channel have similar zero-net-loss current densities, but beyond zero-net-loss, the DP gain increases faster and consistently outperforms the single-pass TE$_0$ gain. From another perspective, the same gain can be achieved with a lower injection current density, or less energy consumption, for the DP-SOA.

**Gain saturation.** Saturation effects of the DP-SOA are measured by inserting sequentially an erbium-doped fiber amplifier (EDFA) and a variable optical attenuator (VOA) between the tunable laser and the input fiber. The measurement schematics can be seen in Supplementary

Fig. S6. The optical power before the VOA is set to 17 dBm (tunable laser + EDFA), while the power at the input fiber is controlled by the VOA. Optical spectra are recorded at the output for each VOA setting, and the optical gain values are obtained by the peak powers subtracting the zero-net-loss levels, same as mentioned above. To calculate the on-chip input power, reference waveguides are measured to calibrate out the FGC loss as mentioned above.

Fig. 4**a** shows the on-chip gain as a function of the on-chip input power for the DP-SOA at a current density of 4.1 kA cm$^{-2}$ and 1570 nm wavelength. The same result for the TE$_0$ channel of the 2 × 2 SOA is also shown as a benchmark. As can be seen, compared to the benchmark, the DP-SOA consistently offers a higher gain of up to 5 dBm on-chip input power. At higher input power saturation occurs. The input saturation power $P_{sat}$ of the DP-SOA is -1.8 dBm, which is lower than that of the single-pass TE$_0$ (6.8 dBm). This is as expected since the DP-SOA has effectively a doubled interaction length. The optical gain increases slightly with the input power before it goes down, which could be due to the beginning and end of the SOA being not well-pumped and acting as saturable absorbers[36]. This can be solved by increasing the pad thickness and making the SOA taper passive.

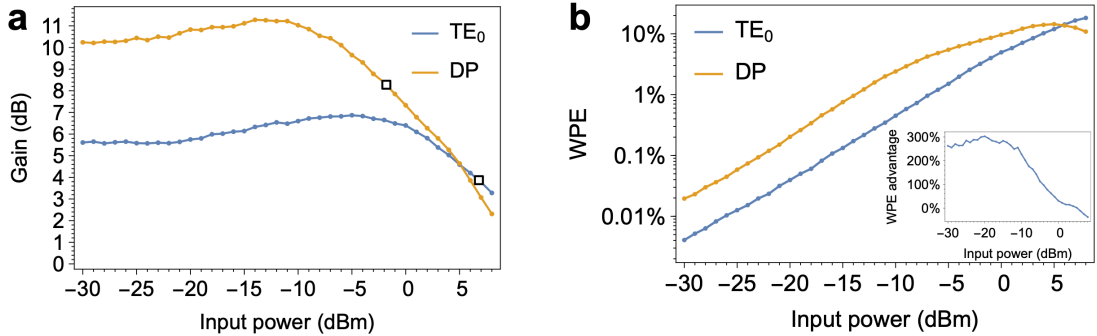

**Figure 4: Saturation effects of the DP-SOA. a, b:** On-chip gain as functions of on-chip input power at the wavelength of 1570 nm, measured with the TE$_0$ channel of the 2 × 2 SOA, and the DP-SOA, respectively. The input saturation points are indicated by empty squares in the figure. **b:** WPE as functions of the on-chip input power, calculated for the TE$_0$ channel of the 2 × 2 SOA, and the DP-SOA, respectively. The inset shows the extra amount of WPE obtained with the DP-SOA.

Fig. 4**b** shows the logarithmic plots of the WPE for both the DP-SOA and single-pass TE$_0$. As can be seen, since the extra gain is obtained without needing more current injection, up to 300% more WPE is obtained for the DP-SOA. The advantage shrinks as the gain becomes saturated, but the DP-SOA remains 100% more efficient until the on-chip input power reaches -5 dBm. At high input powers, the WPE goes down, and this is anticipated to result from increased nonlinear absorption.

**Discussion**

In this study, a 6.2 dB single-pass gain is achieved at 4.1 kA cm$^{-2}$. This value is lower than that of typical bulk InP-based SOAs (~10 dB) with similar active material and dimensions[37], and is attributed to two reasons: the loss in the taper and thermal isolation due to the bonding layer. The passive loss in a single taper is simulated to be around 0.5 dB (Supplementary Note 2), while the absorption is estimated to be 0.8 dB, assuming a 5000 dB cm$^{-1}$ material loss in a single quantum well. Therefore, the single-pass loss, which involves two tapers, is estimated to be around 2.6 dB, meaning an 8.8 dB single-pass gain in the active region in absence of the tapers. This taper loss is comparable to heterogeneously integrated III-V/Si SOAs[38]. The remaining ~1 dB difference is attributed to worse thermal dissipation. It should be noted that neither of these is a fundamental limitation: The taper can be made passive through quantum well intermixing[39], ion implantation, or even be completely eliminated using butt-joint regrowth[40,41]; The thermal performance can be improved by thermal shunting to the silicon substrate[14], or using high thermal-conductivity substrates[16].

Excluding the grating couplers which is not a fundamental limiting factor, the maximum possible gain of the DP-SOA is determined by the crosstalk of the MDM (de)multiplexer, since the crosstalk will manifest itself as reflection into the same mode at the U-bend. In this work, -20.3 dB crosstalk is obtained, but this may be further reduced to -33 dB by altering the waveguide design[42]. This level of reflection is comparable to that in bulk InP SOAs with butt-joint active-passive interface[43].

In conclusion, we demonstrate that gain and efficiency can be significantly enhanced by reinjecting the amplified light for a second pass through the SOA gain section. Each pass in the SOA is in separate transversal modes, so resonance is suppressed and broadband operation is achieved. Light also passes through the gain section from opposite directions, therefore homogeneously utilizing the available carriers. Compared to the single-pass case, the ASE power is improved by > 100%, and 58% – 87% more small-signal gain is obtained at the same injection current in the wavelength range of 1560 nm – 1580 nm. 300% – 100% more WPE is achieved for on-chip input powers up to -5 dBm.

Since the method demonstrated in this paper works at a circuit level, it can be exploited in combination with other efficiency enhancement methods to achieve even better performance. It also allows gain and efficiency enhancement without changing the SOA layer stack itself, providing more compatibility and versatility in circuit design. In principle, the concept is applicable to a wide range of active-passive integration platforms including hybrid, heterogeneous or butt-joint technologies. With the increasingly demanding requirement for BB performance in densely integrated PICs, this work may be used in various applications like optical beam steering, neuromorphic photonics, photonic matrix computation, photonic switching, etc. The DP-SOA may also be used to realize novel laser types. Besides, the design principle can be leveraged to boost the extinction ratio of electro-absorption modulators without increasing the length, so modulation speed will not be compromised. It is also worth noting that this is the first time that mode-division multiplexing is realized in active photonic circuits on-chip, which opens up new opportunities in device and circuit design, such as multi-wavelength lasers or channel multi-casting.

**Methods**

**Fabrication.** The InP epi-wafer (Extended Data Fig. 1**a)** is grown by metalorganic vapor-phase epitaxy (MOVPE). The whole process can be divided into three parts: pre-bonding processing, bonding, and post-bonding processing. In the following process, if not specified, all the patterns in the semiconductor are fabricated with the same procedure: EBL patterning, pattern transfer to a SiN$_x$ hard mask and plasma etching using a CH$_4$/H$_2$ recipe. Pre-bonding (Extended Data Fig. 1**b**): 1) the p-cladding taper is etched together with part of the SOA sidewall. 2) the core taper is etched together with part of the SOA sidewall. 3) the n-cladding taper is etched by a combined dry-wet process, where the dry etch first penetrates into the Q1.25 etch-stop, and then an H$_2$SO$_4$/H$_2$O$_2$ wet etch is used to remove the Q1.25 material completely, without damaging the i-InP waveguide layer. 4) Ni/Ge/Au and Ti/Pt/Au metal layers are deposited on the SOA n- and p-contacts, respectively, using e-beam evaporation. The contacts are then annealed at 400 °C for 30s. Bonding (Extended Data Fig. 1**c**): 1) 100 nm and 50 nm SiO$_2$ layers are deposited on the InP and Si wafers, respectively, for adhesion promotion. 2) the InP wafer is flipped and adhesively bonded with the Si wafer using BCB under vacuum. 3) the InP substrate is removed in an HCl bath, which stops on an InGaAs etch-stop layer. The InGaAs etch-stop is subsequently removed in H$_2$SO$_4$/H$_2$O$_2$, leaving the device layers bonded on Si. The post-bonding process (Extended Data Fig. 1**d**) is as follows: 1) the waveguide (300 nm etch reaching the bonding layer) and grating coupler (120 nm shallow etch) are fabricated, 2) the wafer is planarized using polyimide to fill the 100 nm gap in the MDM (de)multiplexer, 3) the other sidewall of the SOA is etched, forming an S-shaped cross-section, 4) the buried n- and p-metal pads are opened using HCl and H$_2$SO$_4$/H$_2$O$_2$ wet etching, 5) the wafer is planarized again using polyimide, and a final 300 nm-thick pad thickening layer is deposited.

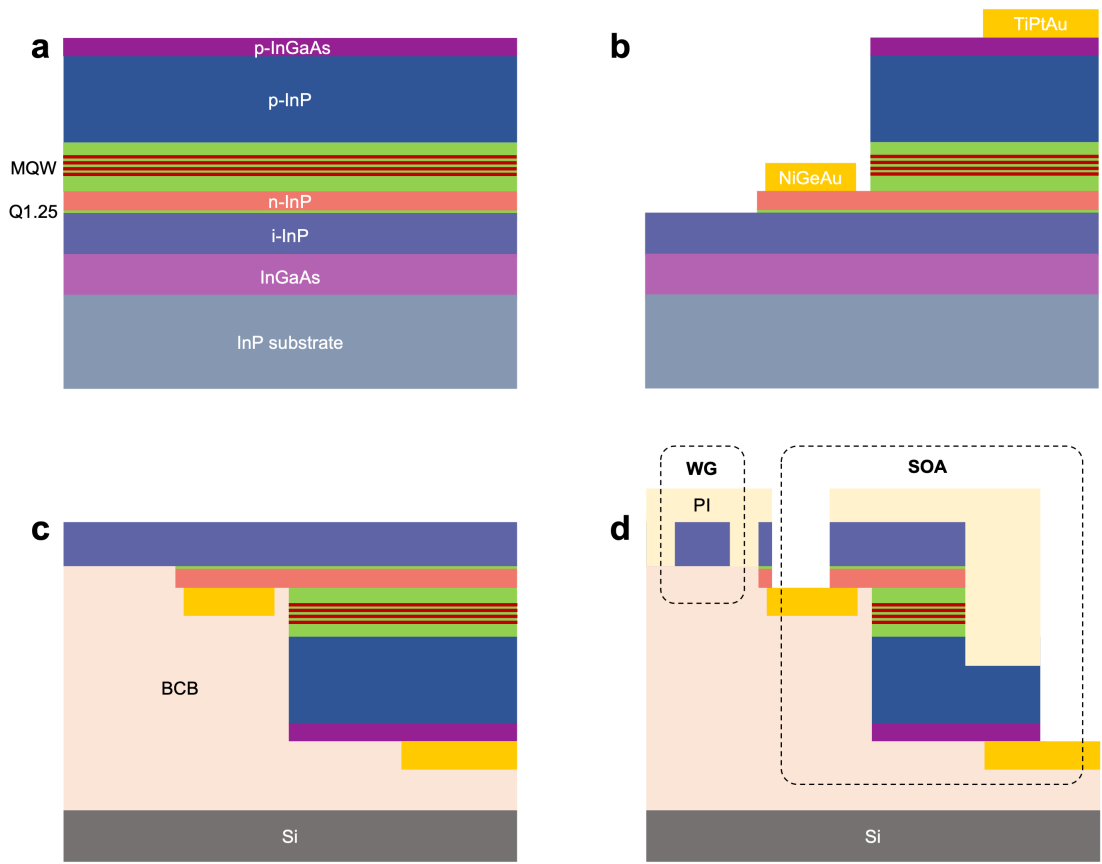

**Extended Data Figure 1: Fabrication of the SOA and passive waveguide. a:** Initial wafer. **b:** Process before bonding. **c:** Bonding. **d:** Process after bonding. Note that the final metal is not shown.

## Acknowledgements

This work is funded by the NWO Zwaartekracht Project "Research Centre for Integrated Nanophotonics", and National Science Fund for Distinguished Young Scholars (61725503) of China. The fabrication was performed in the NanoLab@TU/e cleanroom facility.

## Author contributions

Y. Wang, Y.J., and D.D. proposed the idea. Y. Wang conceived the circuit and device structures. Y. Wei designed the MDM (de)multiplexers. Y. Wang did the simulation, fabrication, and characterization of the circuit. Y. Wang composed the manuscript with input from all co-authors. Y.J. supervised the project. V.D., K.W., M.S., and Y.J. led the overall program.

**Competing financial interests**

The authors declare no competing financial interests.

**Data availability**

The datasets generated during and/or analyzed during the current study are available from the corresponding author upon reasonable request.